\documentclass[aps,prl,superscriptaddress,reprint]{revtex4-2}
\usepackage[english]{babel}
\usepackage{amssymb}
\usepackage{amsmath}
\usepackage[latin9]{inputenc}
\usepackage{array}
\usepackage{multirow}
\usepackage{color}
\usepackage{esint}
\usepackage{bm}
\usepackage{color}
\usepackage{bbm}
\usepackage{hyperref}
\usepackage{babel}
\usepackage{titlesec}
\usepackage{graphicx,amssymb,color}
\usepackage{epsfig}
\usepackage{epstopdf}
\usepackage{lipsum}
\usepackage{mathtools}
\usepackage{soul}
\usepackage[dvipsnames]{xcolor}
\usepackage{tikz}
\usepackage{tcolorbox}
\usepackage{graphics,amssymb,amsmath,epsfig,color}
\usepackage{float}
\usepackage{gensymb}

\begin{document}
\title{Multi-reservoir enhanced loading of tweezer atom arrays}

\author{Xu Yan}
\thanks{These authors contributed equally.}
\affiliation{Department of Physics, The Hong Kong University of Science and Technology, Clear Water Bay, Kowloon, Hong Kong, China} 

\author{Chengdong He}
\thanks{These authors contributed equally.}
\affiliation{Department of Physics, The Hong Kong University of Science and Technology, Clear Water Bay, Kowloon, Hong Kong, China}

\author{Kai Wen}
\affiliation{Microelectronics Thrust, The Hong Kong University of Science and Technology (Guangzhou), Guangzhou, China} 

\author{Zejian Ren}
\affiliation{Microelectronics Thrust, The Hong Kong University of Science and Technology (Guangzhou), Guangzhou, China} 

\author{Preston Tsz Fung Wong}
\affiliation{Department of Physics, The Hong Kong University of Science and Technology, Clear Water Bay, Kowloon, Hong Kong, China}

\author{Elnur Hajiyev}
\affiliation{Department of Physics, The Hong Kong University of Science and Technology, Clear Water Bay, Kowloon, Hong Kong, China}

\author{Gyu-Boong Jo}
\affiliation{Department of Physics, The Hong Kong University of Science and Technology, Clear Water Bay, Kowloon, Hong Kong, China}

\begin{abstract}
We introduce a species-independent method for improved loading into a single-atom optical tweezer array, utilizing iterative loading with multiple reservoir tweezers. Demonstrated with dual wavelength tweezer arrays of $^{88}$Sr atoms, our approach achieves a 96$\%$ loading rate after four reload cycles. This method can significantly enhance existing tweezer rearrangement protocols, potentially reducing iteration time and optical power consumption, thereby enabling a larger number of atoms in a quantum logic device.
\end{abstract}
\maketitle

\paragraph*{\bf Introduction}
Arrays of neutral atoms become a versatile tool for quantum computing, simulation and metrology~\cite{Saffman_2016,Browaeys2020,Kaufman2021,NorciaM.A.2018MCaD,PhysRevLett.122.173201}. With unprecedented control and detection of individual tweezer site, 
a scalable atom array has been demonstrated with realization of several hundreds to a few thousands of qubits~\cite{Ebadi2021,manetsch2024tweezer}, which is usually limited by laser power or field of view. Besides scalability, deterministic preparation of defect-free single atom array is another key ingredient to make quantum logic device based on neutral atom systems.

By finely adjusting the detuning of the light-assisted collision beam, the kinetic energy gained by a pair of atoms through inelastic collision is just enough for one atom to escape from the trap. Near-deterministic loading of single atom arrays~\cite{Grunzweig2010,PhysRevLett.115.073003,FungYH2015Ecbl}, as well as molecules~\cite{WalravenEtienneF2024SfDL}, using this method has been realized. However, the traditional approach to achieve a fully occupied, scalable single-atom tweezers array involves two steps: first, a pairwise loss induced by light-assisted collision to generate a $50\%$ stochastically loaded array, and second, a subsequent rearrangement process to configure the array into the desired setup~\cite{doi:10.1126/science.aah3752,doi:10.1126/science.aah3778,PhysRevApplied.19.034048,KimHyosub2016Issa,Kumar2018,PhysRevLett.122.203601,PhysRevLett.122.143002,PhysRevLett.133.013401,PhysRevX.13.041034,PhysRevA.102.063107,Schlosser_2012}. As the array size increases, both the additional optical power needed for a uniform array and the computation time for the rearrangement signal scale adversely, despite various rearrangement algorithms developed to address this issue~\cite{PhysRevApplied.19.054032,LEE2024150,PhysRevA.108.023107,PhysRevA.108.023108}.

More recently, gray molasses cooling has been combined with the blue-detuned repulsive light-assisted collisions~\cite{WalravenEtienneF2024SfDL,PhysRevX.9.011057,PhysRevX.12.021027,PhysRevLett.115.073003}. This method lower the required trap depth for light-assisted collisions. Some reservoir based deterministic loading method have been demonstrated recently~\cite{PhysRevResearch.5.L032009,gyger2024continuous,norcia2024iterativeassembly171ybatom}. By extracting atoms from a large reservoir and reload to target array many cycles, they achieve over $90\%$ success rate to construct defeat-free atom array. Alternatively, a species-agnostic enhanced-loading method has been demonstrated with bosonic $^{88}$Sr\cite{PhysRevLett.130.193402}. Successfully loaded single-atom tweezers are shielded into dark state, while the empty sites are loaded from the atom source repeatedly. \par

In this work, we present a species-independent enhanced-loading method with dual wavelength tweezers array. Besides the traditional ``target" tweezers where single atom loading involves with light-assisted collisions, another group of far off-resonant ``reservoir" tweezers is generated and loaded with tens to a hundred atoms in each site. Reservoir atoms are moved to an empty target site once detected by fluorescence imaging. Pairwise loss is continuously triggered when multiple atoms move into targets, resulting in $50\%$ filling rate in each reload cycle. 
Our protocol shows a $93\%$ filling rate after three reload cycle from the initial $50\%$ stochastically loaded array, which is consistent with the theoretical $93.75\%$ limit. We also introduce two advanced reload algorithm that enable more than three reload cycles while saving more reservoir power. The filling rate is further improved to $96\%$ with four reload cycles. Additionally, we characterize the cross effect between the cooling beams of dual-wavelength tweezers, which could enable further improvements. Our method illustrate a enhanced-loading algorithm universally work with almost any species of atoms and molecules. It also has unique advantages in saving optical power on target tweezers wavelength, not relying on specific atomic state, and saving iteration time with single magneto-optical trap (MOT) loading process.

\paragraph*{\bf Experiment}

We begin the experiment by capturing atoms in a blue MOT based on $^1S_0$ - $^1P_1$ transition near 461~nm. Repump beams are turned on at the same time to avoid atoms leakage through decay channel $^1P_1$ - $^1D_2$ and then further decay to either metastable states $^3P_0$ or $^3P_2$~\cite{PhysRevResearch.6.013319,JrCWBauschlicher1985Trlo}. Then atoms are transferred to narrow linewidth red MOT near 689~nm, in which atoms are further cooled to a few microkelvin and loading to tweezer traps~\cite{kai2024,ren2024creationtweezerarraycold} (for more details, see Supplementary Materials). 

\begin{figure}[htb!]
	\centering
	\includegraphics[width=9cm]{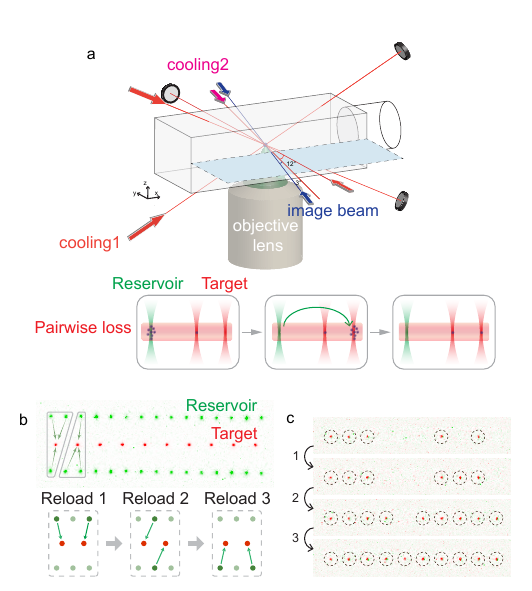}
	\caption{ {\bf Dual wavelength tweezer setup.} (a) The schematic of dual-wavelength tweezers array and reservoir-reload process. Cooling beam 1 cools 813 nm tweezers, while also triggers pairwise loss. Cooling beam 2 takes charge of 515 nm tweezers. Loading multiple atoms from 515 nm reservoirs tweezers. Pairwise loss happens in 813 nm target traps, result in a 50$\%$ reload probability. (b) Overall setups of the reservoir (green) and target (red) tweezers. Preliminary loading algorithm: three reservoirs located around each target in charge of each of the three loadings individually.  (c) A typical reload process with four images, showing the filling rate increased from 50$\%$ to 100$\%$.}
	\label{fig1}
\end{figure}

We prepare two groups of static tweezer arrays with separate Holoeye PLUTO 2.1 series spatial-light modulators (SLMs). The ``target" tweezers lie at the 813~nm magic wavelength of the $^1S_0$ and $^3P_0$ states, where the relative light shift is eliminated. The ``reservoir" tweezers use the high power 515~nm lasers, which also lies around the magic wavelength for $^1S_0$ and $^3P_1$ states. All of the tweezers are focused by a NA=0.5 microscope objective lens at the atom plane (Fig.~\ref{fig1}a), resulting in a $\simeq$0.7$\mu m$ beam waist tweezer traps for 813~nm. 

\begin{figure*}[htb!]
	\centering
	\includegraphics[width=16cm]{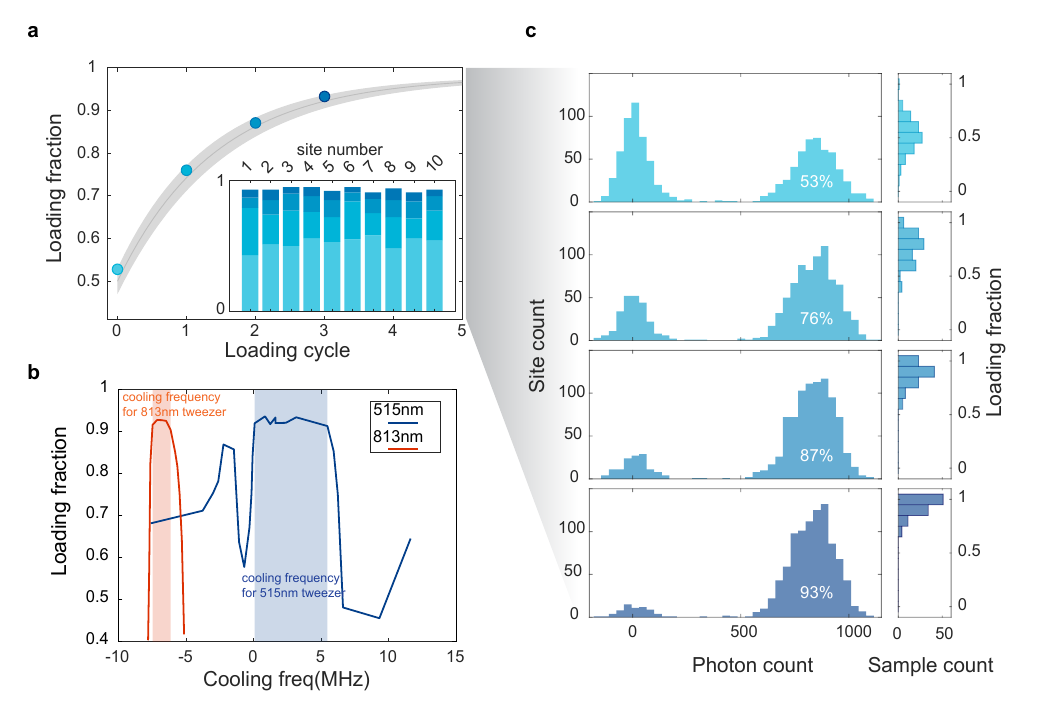}
	\caption{ {\bf Multi-reservoir enhanced loading.} (a) Filling fraction of tweezer sites increases after reloading cycle, the filling fraction for initial loading and subsequent three reloading cycles is 0.53,0.76,0.87,0.93 correspondingly. Shaded area shows expected value of filling fraction considering initial loading fraction usually varies from 0.47 to 0.53. The inset figure shows detailed reloading process for all ten sites. (b) Final filling fraction depends on both 515nm and 813nm tweezer cooling beam detuning. Shaded area shows good working frequency range of both cooling beam detuning, there is no overlap between them. (c) Left column if the histogram of fluorescence photon counts from all tweezers site with 100 repeat measurements. The right column shows histogram of filling fraction for those 100 samples.}
	\label{fig2}
\end{figure*}

Our tweezer configuration includes one line of 10 target tweezers in the middle, with 30 reservoirs on top and bottom as illustrated in Fig.\ref{fig1}b. In the initial setup, we assign three specific reservoirs (see triangle containing three reservoir tweezer sites in Fig.~\ref{fig1}b) to each target tweezer site, corresponding to three reload cycles. Both groups of tweezers are risen up to 100~$\mu$K during the red MOT stage, atoms are only loaded in the focus plane of those shallow tweezer sites. We further compress red MOT within 90~ms to increase atom density and  ensure all tweezers are loaded with atoms. Then we turn off red MOT beam and hold atoms with tweezer traps for another 40~ms. During this holding stage, we adjust bias coil current to compensate residual B field to zero and rise 813~nm tweezer to 1.0~mK and 515~nm tweezer to 1.5~mK. Cooling beam for both tweezer arrays are turned on. The cooling of 813~nm tweezers also triggers light-assisted collision and pairwise loss, resulting in a $50\%$ initial loading rate in the targets.

We setup another 515~nm dynamic tweezer to move atoms between reservoirs and targets (A schematic of beam path can be found in Supplementary Materials). The movable tweezer is dynamically controlled by an AA opto-electronics DTSXY-400 biaxial acousto-optical deflector (AOD), which is driven by RF signals from a Spectrum m4i.6622-x8 arbitrary waveform generator (AWG). We pre-measure all the RF-frequencies that overlap the AOD tweezers to each target and reservoir sites by absorption imaging. To reduce experiment runtime, we pre-calculate all potential RF waveforms and stored them in the computer memory. 

\begin{figure*}[htb!]
	\centering
	\includegraphics[width=16cm]{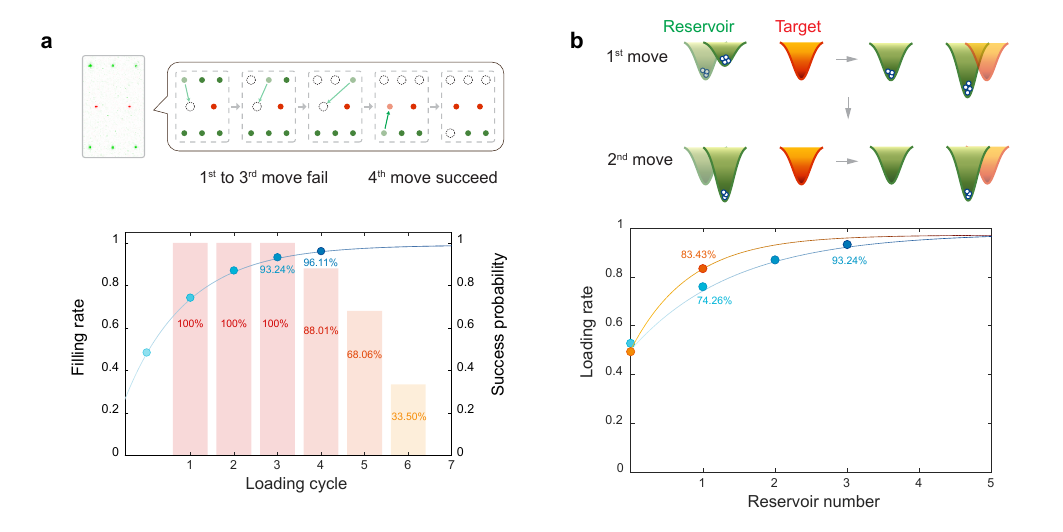}
	\caption{ {\bf Advanced loading algorithms with improved loading performance.} (a) Sharing reservoirs between two target traps. At each loading, each empty trap searches the available reservoirs from either end. A 96$\%$ filling rate is recorded after 4 loadings under the new algorithm. Red bars shows the chance to have adequate reservoirs at each loading cycle.  (b) Reuse each reservoir for two reload process. Reduce the initial trap depth of first reload to move only part of the reservoir atoms. The trap depth resumes during movement to prevent atom loss. Second reload holds maximum trap depth to move all remaining atoms from reservoir. An 83$\%$ filling rate is recorded with reusing single reservoir twice. }
	\label{fig3}
\end{figure*}

\vspace{10pt}

\paragraph*{\bf Multi-reservoir enhanced loading} We first take the fluorescence image with 461~nm $^1$S$_0$ to $^1$P$_1$ transition.  When we turn on image beams, repump beams are turned on simultaneously to prevent atoms leakage~\cite{PhysRevLett.133.013401}. It is worth to notice that the differential AC stark shift for $^1P_1$ relative to $^1S_0$ state in 515~nm and 813~nm tweezer are opposite. In the 813~nm trap, $^1S_0$ - $^1P_1$ resonance frequency shift to red side, while in 515~nm trap resonance frequency shift to blue side. This feature enables us to minimize image heating effect in reservoir tweezers since fluorescence beams are off-resonance for atoms in the 515~nm trap.  

We begine reloading cycle after first fluorescence image. Each reloading cycle starts with analyzing loading results from previous cycle. The occupation status of the targets sites is extracted and sent to the AWG control module. For those empty target sites, we move atoms from one designated reservoir site to reload them. Desired RF signals are selected, combined and sent to the AOD to control the reload beams. Multiple atoms are moved from reservoirs to those unoccupied targets. The cooling beam is kept on to continuously trigger pairwise loss in target sites, so that the ideal reload probability will be $50\%$ for each single reload. We repeat this cycle to get the final filling rate near $100\%$.

We show the result of a typical loading process for 10 target tweezer sites in Fig.\ref{fig2}a.
In this sample, the initial filling fraction is $53\%$, after the first reloading cycle, the filling fraction is $76\%$ and after the third reloading cycle, the filling fraction becomes $93\%$. And up to the third reloading cycle, all the sites has expected filling fraction over $90\%$ as shown by inset figure in Fig.\ref{fig2}a.
We can use a simple model to describe our reloading process. Assuming our initial loading probability is $f_0$, for each reloading cycle the reloading probability for empty states is $d$, atom survival rate during each cycle is $p$. Then the filling fraction after $n$th reloading cycle is given by recurrence relation $f_n = [(1-f_{n-1})d+f_{n-1}]p$. Eventually, the filling fraction can be expressed by
\begin{equation}
	f_n = pd\frac{1-p^n(1-d)^n}{1-p(1-d)}+p^n(1-d)^nf_0
	\label{refill_eq}
\end{equation}

Considering the initial loading fraction $f_0$ usually varies from $47\%$ to $53\%$, reloading probability $d$ is $50\%$, cycle-to-cycle survivability $p$ is $99\%$. We plot a expect value of reloading fraction as a shaded area in Fig.\ref{fig2}a, which matches our data near perfectly. The histogram of fluorescence counts from tweezer sites can also reflect our reloading process. As shown in Fig.\ref{fig2}c, left peak represent those sites without atom and right peak represent those with single atom. In the beginning, they are balanced, $50\%$ for both peaks. As we reload empty sites in each reloading cycle, the right peak grows and left peak shrinks down. 

The cooling for 813~nm target tweezers array is attractive Sisyphus cooling, in which the excited state experiencing a deeper trapping potential than the ground state. In contrast to the target tweezers, cooling scheme of our 515~nm reservoir tweezers is repulsive Sisyphus cooling. Even though, frequency of both cooling beams close to the same narrow line transition $^1S_0$ - $^3P_1$, the optimal detuning are quite different. As shown in Fig.\ref{fig2}b, we measured the final reloading fraction after 3 reloading cycle versus cooling beam detuning. For 813nm target array, the optimal cooling frequency is red detuned around 6.8MHz, while for 515nm reservoir array the optimal cooling frequency is blue detuned around 1.6MHz. There is heating effect for atoms in 515nm tweezers array caused by 813nm array cooling beam, that cause atom loss in reservoir array. In this case, using multi-atoms reservoir array is more robust way, as long as reservoir tweezer is not empty, $50\%$ reloading probability is guaranteed. Another side, cooling beam for 515nm tweezers also cause certain heating effect for atoms in 813nm tweezers, which induce around $1\%$ atom loss in each reloading cycle.

\paragraph*{\bf Advanced reservoir-enhanced loading} In the general reservoir-reload process, we noticed that the reload probability of single move is limited to $50\%$ by the pairwise loss mechanism. To reach higher filling rate, we have to perform more potential moves with a fixed reservoir number and laser power. Here we further introduce two advanced reload algorithms to go beyond the $93.75\%$ limit. 
    
The first advanced algorithm aims to detect and make full use of spare reservoirs. In the original algorithm, there is inevitable waste of unused reservoirs. We assigned a fixed number of reservoirs to each target, and each reservoir in charge of the movement in a specific reload cycle. If a target is already filled after certain cycle, its remaining reservoirs will be left unused. We attempt to use these targets
As shown in Fig.~\ref{fig3}a, we start with sharing six reservoirs between two adjacent targets. While one of the targets is occupied, all the remaining reservoirs can be used for extra reload cycles on the empty sites. The schematic shows a typical case, where the one of the two-target group is filled, while the other is empty. All six reservoirs can be used for the empty sites, instead of three in the original method. Fig.~\ref{fig3}b shows that when sharing reservoirs between two targets, the fourth reload mostly happens successfully with a $96\%$ filling rate. 
We also track the ``success rate" under this algorithm: a reload is considered failed if no available reservoir can be found when an empty target needs to be filled. Obviously, the original algorithm always succeed for first three moves, and fail for further ones. Simulation shows that under this improved algorithm, the fourth still keeps a $90\%$ success rate. Statistically, more than 4 reloads is still possible. But when sharing reservoirs between only two targets, there is a high chance to have inadequate available reservoir starting from the 5th reload, as shown in Fig.\ref{fig3}a. 
Simulation shows that as we share more reservoirs among more targets, the success probability of 4th reloading finally increases to 1. It also shows that if we share all the reservoirs among all the targets, less reservoirs would be needed for fixed number of reloading cycle.

Another approach is to reuse a single reservoir more than once. 
In the default experiment setup, the moving tweezers is 1.5 times as deep as the reservoirs. It moves all the atoms out of reservoir at once, deeper trap also reduces atom loss during the moving. 
To reuse each reservoir, we reduce the moving tweezers power to make it slightly shallower than the reservoirs. As shown in Fig.\ref{fig3}b, about half of the reservoir atoms remain in the trap, while the other half are moved. The cost is considerable atom loss through moving process, which reduces the reload probability of the first move to $~30\%$ from ideal $50\%$. 
The remaining atoms will be used for a second move. For the second move of same reservoir, we resume the moving tweezers power to move all remaining atoms, and then perform the standard imaging-reloading procedure. A significant increase to $83\%$ refill rate is recorded with two moves from single reservoir, compared to the $75\%$ from single move. Fig.\ref{fig3}b shows the expected reload rate versus reservoir number, where it reaches more than $95\%$ with three reservoirs.

\begin{figure}
	\centering
	\includegraphics[width=9cm]{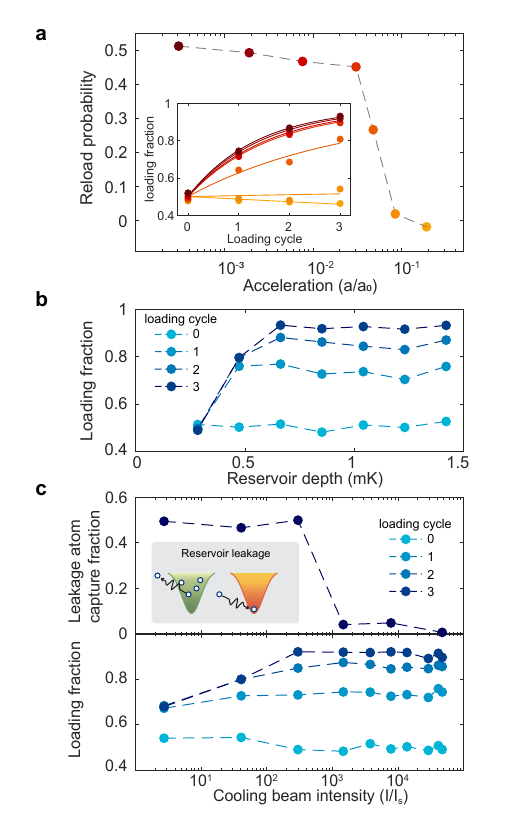}
	\caption{{\bf Optimal reloading conditions} (a) We move atoms from reservoir to target sites using AOD tweezer with  constant acceleration and deceleration. To ensure $50\%$ reloading rate, we need to make the acceleration is much smaller than $a_0$. The inset shows performance through three cycles under different accelerations. (b) Loading fraction dependence on reservoir depth. Atoms in deeper reservoirs can easily survive several reloading cycles. (c) Target tweezer loading fraction dependence on reservoir tweezers cooling beam intensity. When the cooling beam intensity is not strong enough, there will be leakage atoms from reservoir and those atoms could collide with atoms in target tweezers and induce atom loss.}
	\label{fig4}
\end{figure}

\paragraph*{\bf Optimal reloading conditions} Finally, we calibrated the optimal reloading conditions based on final loading rate. We move atoms from reservoir sites to target sites with constant acceleration and ~\cite{Hwang:23}. In an accelerated moving optical trap form by Gaussian beams, the effective trap depth decreases due to acceleration. The trap local minimum point will disappear when keep increasing acceleration to $a_0$. This critical acceleration $a_0 \simeq 327,000 m/s^2$ is calculated based on our trap parameters. By testing varies accelerations, we found to keep reload probability close to ideal value $50\%$, the accelerations should be smaller than 0.05$a_0$ as shown in Fig.~\ref{fig4}a.
Reservoir trap should be deep enough to hold atoms with heating effect from imaging beam pulse and cooling beam for target sites. We plot target sites loading fraction while varying reservoir trap depth in Fig.~\ref{fig4}b. To maintain a high final loading fraction, the lowest reservoir depth is 0.7~mK.
Atoms in reservoir traps must be cooled down with cooling beam, otherwise there will be obvious leakage effect. Atoms escaped from reservoir has the chance to be captured by target site. We observed this effect when cooling beam intensity is not strong enough as shown in Fig.~\ref{fig4}c. When there is one atom in a target site, two atoms will collide with each other assisted by pair loss beam, and loading fraction drops. Limited by multi atoms life time in reservoir traps, each loading cycle should be finished within 2s to keep high final filling fraction.

\paragraph*{\bf Conclusion}
In summary, we have demonstrated a near deterministic enhanced loading of single atom array by reloading it for a few cycles with multiple reservoirs. The filling fraction is boosted up to $96\%$ within 4 reloading cycles. Reservoir array pattern can be adjusted according to target array geometry to maintain relative short transport distance and high reloading efficiency. The method we showed here will not be limited by target array size and atom species. Moreover, by choosing wavelength of reservoir tweezer to make both array have consistent cooling condition, single atom reservoir can be realized, which would enable a deterministic reloading of target array within one reloading cycle. Deterministically filling of larger arrays will accelerate the study related to metrology, quantum simulation, quantum calculations with optical tweezer systems.

\vspace{10pt}
\paragraph*{\bf Acknowledgement}
GBJ acknowledges support from the RGC through 16302420, 16302821, 16306321, 16306922, 16302123, C6009-20G, N-HKUST636-22, and RFS2122-6S04. CH acknowledges support from the RGC for RGC Postdoctoral fellowship.

\vspace{10pt}
Corresponding author email: \\cheab@connect.ust.hk; gbjo@ust.hk
\selectlanguage{english}
\bibliography{enhanced_loading}


\clearpage

\onecolumngrid
\begin{center}
	\textbf{\Large Supplemental Materials\\ for\\"Multi-reservoir enhanced loading of tweezer atom arrays"}
\end{center}

\setcounter{equation}{0}
\setcounter{figure}{0}
\setcounter{table}{0}
\setcounter{page}{1}
\makeatletter
\renewcommand{\theequation}{S\arabic{equation}}
\renewcommand{\thefigure}{S\arabic{figure}}
\renewcommand{\bibnumfmt}[1]{[S#1]}
\renewcommand{\citenumfont}[1]{S#1}

\begin{figure}[!b]
	\centering
	\includegraphics[width=8cm]{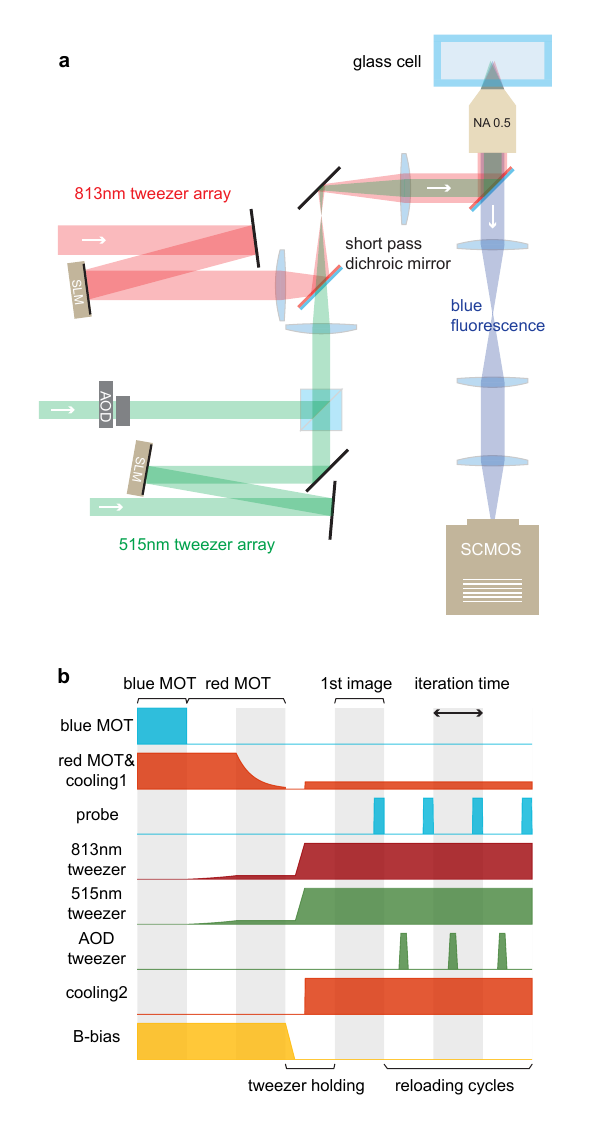}
	\caption{\textbf{Experimental details}(a) Experimental setups for dual-wavelength tweezer array, microscope objective and fluorescence imaging. Dichroic mirrors are used to combine tweezers and separate fluorescence.  (b) Experimental sequence for enhanced loading cycle, from MOT cooling to three consecutive reload cycles. }
	\label{layout}
\end{figure}

The setup of dual-wavelength tweezers and imaging system are shown in Fig.~\ref{layout}a. Both Spatial light modulator (SLM) paths have $96\%$ reflection efficiency under $5\degree$ incident angle. The AOD path has a $60\%$ efficiency. The two 515~nm tweezer beams are combined by a polarizing beam splitter (PBS), then coupled with a 813~nm beams by a dichroic mirror. Another dichroic mirror locates just at the entrance of the microscope objective to separate the tweezers and blue fluorescence. We image from the same objective to help alignment.

The detailed experimental sequence is shown in Fig.~\ref{layout}b. We start with a broad 461 nm MOT to capture hot Sr atoms for 1 second. Two repump beams at 679 nm and 707 nm are kept on to repump metastable state atoms back to the cooling cycle. The atoms are then transferred to a broadband red MOT for 90 ms, during which it is gradually compressed and further cooled to $2\mu$K. Dual-wavelength tweezers are turned on to a low power during red MOT. Tweezers powers are raised to working value after MOT is turned off. We capture the blue fluorescence of tweezers atoms by a Tucsen Dhyana 400 BSI V3 sCMOS camera. Cooling beams for both wavelengths are kept on during following tweezers stage. We iteratively image the tweezers atoms, analyze site filling status and perform reservoir reload. The iteration time is defined by the time spent on image analysis and RF signal computation.

\begin{figure}
	\centering
	\includegraphics[width=9cm]{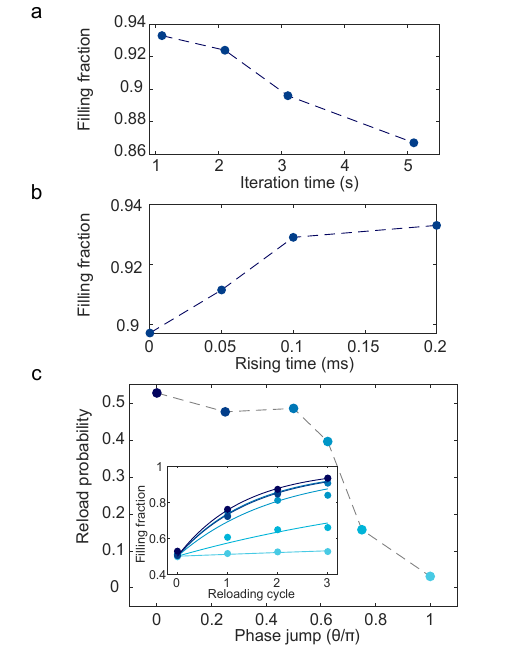}
	\caption{\textbf{Optimize reloading parameters} (a)Due to limited lifetime for atom in tweezers, final filling fraction drops if we spend too long time in each cycle. (b) When we moving atom from reservoir to target sites, ramp up AOD tweeezer power instead of suddenly switch on will help to increase filling fraction. (c) Phase of rf signal for controlling AOD should be continued in order to successfully transport atom from reservoir to target sites.}
	\label{sub1}
\end{figure}

Due to the limited lifetime of atoms in reservoir tweezer traps, the longer time we spend in each reloading cycle, the lower final filling fraction we can achieve. As shown in Fig.~\ref{sub1}a, when we spend more than 3s in each reloading cycle, the filling fraction drop to below $90\%$. It's better to know the limitation of our method even we only need less than 1s in each cycle. To ensure successful transport of atom from reservoirs to targets, we first ramp up moving AOD tweezer slowly enough and then move it as smooth as we can. If ramp up process is too fast, the final filling fraction will drop as demonstrated in Fig.~\ref{sub1}b. Smoothly moving AOD tweezer is important. If there is a phase jump in rf signal, which driving AOD, serious atom loss could happen as shown in Fig.~\ref{sub1}c.

\begin{figure}
	\centering
	\includegraphics[width=9cm]{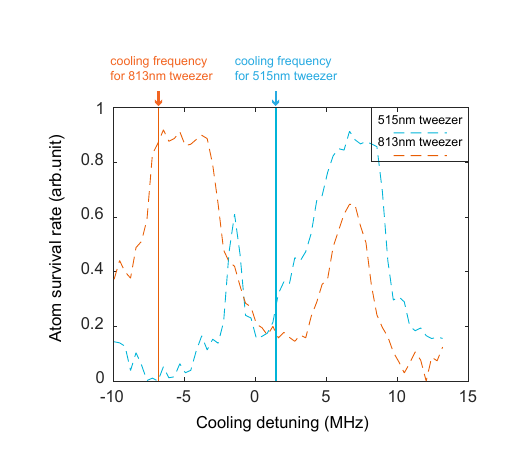}
	\caption{\textbf{Atom loss induced by cooling beam.} The cooling beam frequency for reservoir and target tweezer array are different. There is heating effect in 813~nm tweezer from cooling beam for 515~nm tweezer, which contribute to additional atom loss about $1\%$ for each reloading cycle.}
	\label{crosstalk}
\end{figure}

As we showed in main text, the cooling beam for both reservoirs and targets are necessary. The cooling beam frequency for two tweezer array are different, which leads to a cross talk between them. The cooling beam for 515nm tweezer has heating effect for 813nm tweezer and vice versa. As shown in Fig.~\ref{crosstalk}, the optimal choices of cooling beam frequency for both tweezers not only efficiently cool down atoms but also minimize cross talk between two tweezer array.

\end{document}